# Amplification of Molecular Chiroptical Effect by Low-loss Dielectric Nanoantennas

*Weixuan Zhang, Tong Wu, Rongyao Wang and Xiangdong Zhang\**

Beijing Key Laboratory of Nanophotonics & Ultrafine Optoelectronic Systems, School of Physics, Beijing Institute of Technology, Beijing, 100081, China

**ABSTRACT:** We report here chiroptical amplification effect occurring in the hybrid systems consisting of chiral molecules and Si nanostructures. Under resonant excitation of circularly polarized light, the hybrid systems show strong CD induction signals at optical frequency, which arise from both the electric and magnetic responses of the Si nanostructures. More interestingly, the induced CD signals from Si-based dielectric nanoantennas are always larger than that from Au-based plasmonic counterparts. The related physical origin was disclosed. Furthermore, compared to the Au-based high-loss plasmonic nanoantenas, Si-based low-loss structures would generate negligible photothermal effect, which makes Si nanoantennas become an optimized candidate to amplify molecular CD signals with ultralow thermal damages. Our findings may provide a guideline for the design of novel chiral nanosensors for applications in the fields of biomedicine and pharmaceutics.

**Introduction**

Chirality plays a crucial role in modern biochemistry and the evolution of life.[1,2] Many biologically active molecules are chiral, detecting and characterizing chiral enantiomers of these biomolecules are of considerable importance for biomedical diagnostics and pathogen analyses.[3] A common technique for chirality detection/charaterization is circular dichroism (CD) spectroscopy describing the difference in molecular absorption of left- and right- handed circularly polarized photons.[4,5] Recently, intense theoretical[6-13] and experimental[14-20] studies have been devoted to the plasmon-enhanced CD spectroscopy, with an aim of realizing ultrasensitive chirality detections by means of plasmon-enhanced light-matter interactions. For some specially designed plasmonic nanostructures, such as Metal−Insulator−Metal dimers,[21] metallic split-ring resonators,[22] nanocups[23] and double fishnet structures,[24] early works revealed that both electric and magnetic responses at the optical frequency[21-24] could play pivotal role in generation of chiral near fields with a specific handedness.[25, 26] Such chiral near fields could tailor the spontaneous decay rate of chiral molecules[24] that are located in the surrounding areas of the plasmonic nanostructures. In this way, weak optical activities of molecules (normally in UV spectral region) could be both enhanced and transferred to the plasmonic absorptions of metallic nanoparticles (in Vis/NIR region), thus enabling a highly sensitive way for chiroptical detection of biomolecules. Particularly, the induced plasmonic CD spectroscopy has been used successfully for detection of amino acids,[14] peptides,[17] and DNA[18] showing a much higher sensitivity than that of the conventional CD technique.

Despite of these progresses, applications of plasmonic CD technique suffer from high optical losses caused by the resonant excitation of metallic nanoparticles at optical frequency. It has been found that the high optical losses can induce huge photothermal effect, which caused changes in dielectric properties of the nanoparticles and surrounding medium, and even a modification of adsorption/desorption kinetics of molecules at metallic surfaces.[27-33] In addition, a previous research has shown that heat treatment of $\beta-Lactoglobulin\,(\beta-Lg)$ results in at least two irreversible changes to the secondary structure of $\beta-Lg$.[34] Polyethylene glycol thiol will undergo the irreversible transformation from the helical state to an elongated form in a high temperature environment.[35] These structural changes can alter the chirality of protein and make the CD spectrum become useless. Thus low-loss nanoantenas are required to overcome these photothermal problems.

On the other hand, the Mie-type responses in high-permittivity nanoparticles, which show extremely low optical losses,[36-39] have been a subject of particular interest in recent years. These dielectric nanoparticles, such as the silicon nanosphere, can be utilized in direct analogy to the plasmonic resonances of metallic nanoparticles to engineer their optical response towards specific applications.[40-57] For example, the interference between the overlapped electric and magnetic dipolar modes allows to fulfill a condition for almost zero backward light scattering.[40-43] The low intrinsic losses dielectric nanostructures can sustain much higher optical powers and ultimately provide orders of magnitude higher frequency conversion efficiency for the study of nonlinear effects.[44-49] The high efficiency photonic metasurfaces based on the high-index nanoresonators have been designed to manipulate phases and polarizations of light waves.[50-52] The dielectric nanoparticles have also been used for unidirectional excitation of electromagnetic guided modes[53], enhancing the Raman scattering[54], probing of the surface-enhanced Raman optical activity of chiral molecules[55] and tailoring the chirality of light emission.[56] Furthermore, the enhanced enantioselective absorption caused by molecular vibrational modes is achieved using Si submicrometer sphere[57]. However, whether low-loss dielectric nanoparticles can be used to induce stronger surface-enhanced CD spectroscopy (both of molecular and nanostructure-induced CD) in visible region has not been addressed in previous studies.

In this work, we discuss the chiroptical effect occurring in the hybrid systems consisting of chiral molecules and Si nanostructures by means of T-matrix and finite element methods. Taking nanospheres and periodic nanoribbons as model systems, we demonstrate that the stronger CD spectroscopy can be produced with the combined exciting of both electric and magnetic resonances in the dielectric nanostructure, which can boost the interactions between chiral molecules and nanoantennas. In addition, the silicon nanoantenna related heating is much lower than that of the Au plasmon-analogue system, which is beneficial to both photophysical and chemical processes for chirality discrimination.

**Surface-enhanced CD spectra based on the single nanoantenna systems**

In this section, we will discuss the surface-enhanced CD based on the single nanoantenna systems. Firstly, we consider a hybrid system consisting of a chiral molecule and a nanosphere (NS), as shown in Fig. 1a, which is excited by circularly polarized plane waves with the angular frequency $\omega$. The chiral molecule is put at the origin of the coordinate, and the NS with radius $R$ locates at $R_s=(0,0,R+d)$. Here $d$ is the distance between the chiral molecule and the surface of the NS. The chiral molecule used here is assumed as a combination of the molecular electric dipole (**u**) and magnetic dipole (**m**). The corresponding parameters used for **u** and **m** are given by: **u**$=(0,0,|e|r_1)$ and **m**$=(0,0,0.5i|e|r_0r_1w_0)$ with $r_1$=0.2 nm, $r_0$=0.005 nm and the wavelength of the molecular transition being $\lambda_0 = 2\pi c/\omega_0 = 300\,\text{nm}$.[7,8] Using the T-matrix method[7] and assuming that the hybrids dispersed in water are randomly oriented, we can calculate the CD spectrum of the system (averaging the solid angle for the incident wave vector in the calculation). The CD signals calculated for the NS systems are the differences of the extinction efficiency between the left- and right-hand circularly polarized excitations of the hybrids. For comparisons, both Au and Si NSs with same radius ($R$=65 nm) are considered. For the dielectric functions of silicon and gold, Johnson's data are adopted.[58] The refractive index of water is 1.33.

As shown in Fig. 1b, we can see that the CD peak intensity produced by the Si NS (black line) is four times larger than that by the Au NS (red line). To disclose its physical origin, we plot the enhancement of the optical chirality density at the position of the single chiral molecule. The optical chirality density can be expressed as: $C = \varepsilon_0 \vec{E}\cdot\nabla\times\vec{E}/2 + \vec{B}\cdot\nabla\times\vec{B}/2\mu_0$, where $\varepsilon_0$ and $\mu_0$ are the permittivity and permeability of free space, respectively, $\vec{E}$ and $\vec{B}$ are the local electric and magnetic fields.[59,60] According to the previous studies,[61] the amplification of the optical chirality density on the molecular position can enhance the molecule-antenna chiral interaction and result in a stronger plasmonic CD spectroscopy. As shown in Fig. 1c, the Si NS can produce a larger enhancement of the optical chirality density ($C/C_0$) at the chiral molecular position, which will lead to a more significant CD effect. Herein $C_0$ is the optical chirality density for the incident left-hand circular polarized light. Fig. 1d presents the comparison of the far-field extinction efficiency between Au and Si NSs ($R$=65 nm). Three peaks of the Si-based extinction efficiency (black line) can be clearly identified for the incident wavelengths being 551nm, 476nm and 389nm, which correspond to the magnetic dipolar, electric dipolar, and magnetic quadrupolar resonances, respectively. In contrast, the Au NS only sustains electric dipolar resonance (red line). When the Au nanoparticle is on the electric resonance, the phase of the scattered electric field is nearly $0.5\pi$ delayed with respect to the magnetic field. Therefore, the electromagnetic density of chirality ($C\sim\text{Im}(\boldsymbol{E}^*\cdot\boldsymbol{B})$) remains weak, although the metallic nanostructures may process strong electric resonances. Combined exciting of both electric and magnetic resonances in Si nanostructures, the ideal phase relation between the electric and magnetic fields can be produced at the molecular position, which may lead to a large electromagnetic density of chirality.[26]

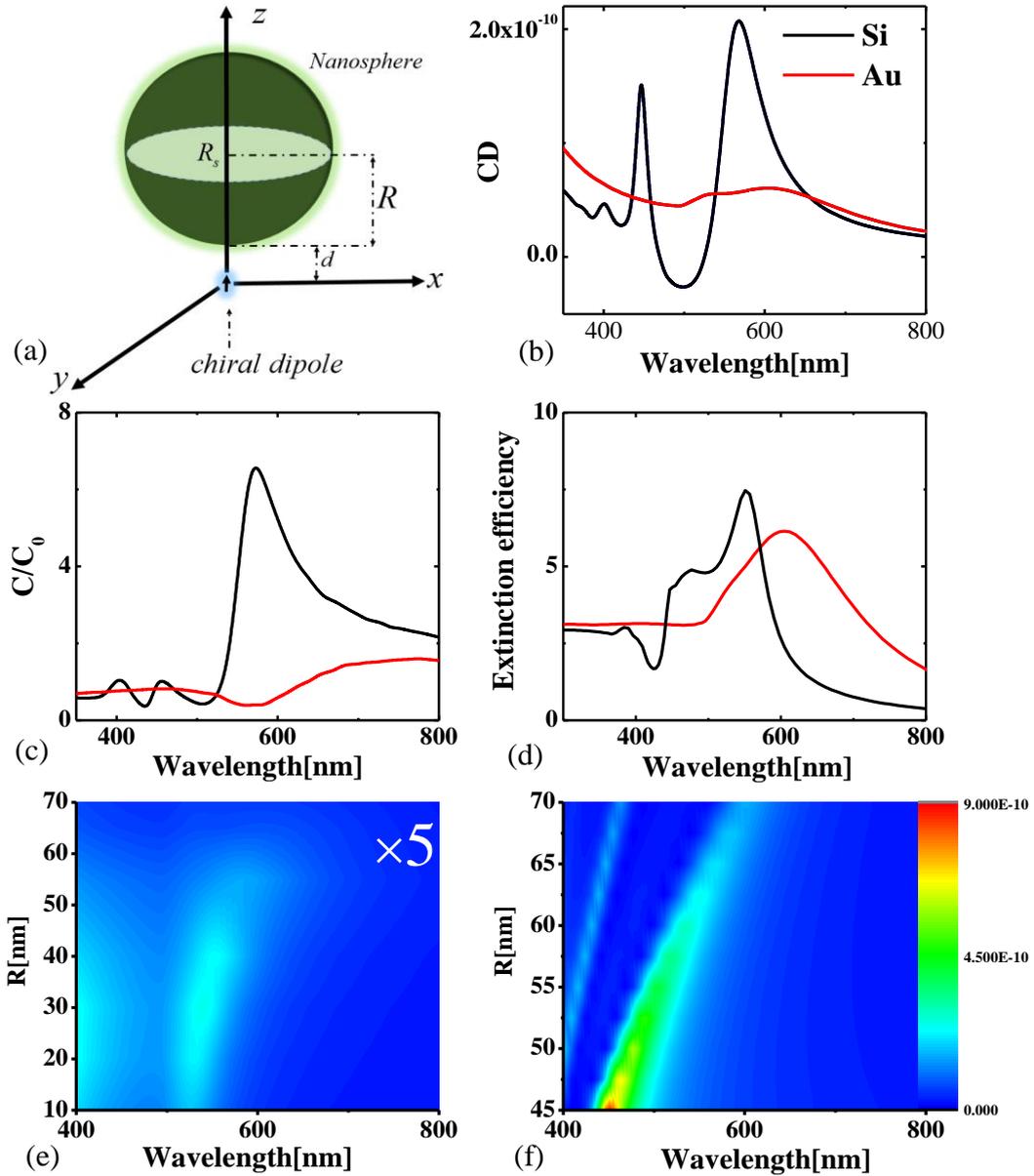

**Fig. 1** (a) shows the geometry and coordinate of the hybrid system, which consists of a chiral molecule and a NS. The chiral molecule is put at the origin of the coordinate. The NS, with radius $R$, locates at $R_s$. (b) The spectra for the enhanced CD single based on the Si (black line) and Au (red line) NSs. The radius of the NS is $R$=65 nm. The distance between the molecule and the surface of the NS is $d$=1 nm. (c) shows the enhancement of the electromagnetic density of chirality as a function of the incident wavelength for the single NS. (d) shows the far-field extinction efficiency for the single NS. (e) and (f) display the influence on the absolute values of enhanced CD signals by changing the radii of the Au (×5) and Si NSs, respectively.

In order to investigate the influence of the nanoparticles' sizes on the surface-enhanced CD spectra, we calculate the corresponding CD signals induced by the Au and Si NSs with various radii. The results are shown in Fig. 1e (for Au NSs) and 1f (for Si NSs). Here the CD signals induced by the Au NSs are multiplied by an amplification factor 5. We find that the optimized radius of the Au NS to enhance the CD signal is nearly R=30nm. However, the corresponding CD peak is still much lower than the Si-based (R=40~70nm) CD signals. With the decrease of the radii of Si NSs, the CD signals increase due to the blue-shift of the resonance modes (closing to the wavelength of the molecular transition).

The above results only focus on enhanced CD spectra of the hybrid system comprised of a single chiral molecule defined as a point dipole and an isolated NS. In fact, there are many theoretical and experimental investigates about chiral detection using a periodic plasmonic metasurface platform.[12, 14, 15] Consequently, it is necessary to study the CD spectra of the hybrid systems composed of chiral molecule media and periodic nanostructures. The corresponding schematic is shown in Fig. 2a, which is easily fabricated in real experiments. The Si nanoribbon (NR) (infinite along y-axis) processes a square cross-section in x-z plane with the length $D=120$ nm. The chiral medium strips, which cling to both sides of the NR, are infinite in y direction, 5 nm and 120 nm along x- and z-axis, respectively. The molecule-NR hybrids are arranged periodically. Based on the numerical validation, we find that the influences of higher diffraction orders on the CD signal are negligible with the period being 350nm.[66] In our studies of the periodic medium-NR nanocomposites, we chose to modify a standard COMSOL RF simulation environment by implementing chiral constitutive relations:[62-66]

$$\boldsymbol{D} = \varepsilon_0 \tilde{\varepsilon} \boldsymbol{E} + \tilde{\kappa} \boldsymbol{B}$$
$$\boldsymbol{B} = \mu_0 (\boldsymbol{H} - \tilde{\kappa}^T \boldsymbol{E}),$$
(1)

where the superscript '$T$' indicates transposition; $\varepsilon_0$ and $\mu_0$ denote the vacuum permittivity and permeability, respectively. The chirality parameter $\tilde{\kappa}$ and dielectric tensors $\tilde{\varepsilon}$ for the oriented chiral molecular medium are derived from the first-principles calculations.[66] The molecular density of the chiral medium used in our calculations is $n_0=(2\text{ nm})^{-3}$ and the corresponding molecular electric and magnetic dipoles are expressed as: $\boldsymbol{u}=(|e|r_1,0,0)$ and $\boldsymbol{m}=(0.5i|e|r_0 r_1 w_0,0,0)$. The CD signals calculated for the periodic medium-NR nanocomposites are defined as the differences of the transmittance spectra between the left- and right-hand circularly polarized excitations.

As shown in Fig. 2b, the CD signal induced by the periodic Si NRs is much larger than that by the Au counterpart. In Fig. 2c, we present the comparison of the averaged optical chirality density on the positions of the chiral medium strips induced by Au and Si NRs, respectively. The averaged optical chirality density can be expressed as: $C_A = \int_{V_{medium}} C(x,y)/C_0 dV$ with $V_{medium}$ being the region occupied by the chiral medium strips. It is clearly shown that the periodic Si NRs can produce a larger $C_A$ than Au NRs on the positions of the chiral media, which is due to the fact that the Si NR possesses both electric and magnetic resonances. This can be clearly seen in the transmission spectra of the periodic Si and Au NRs in Fig. 2d. Three peaks (630nm, 490nm and 460nm) of the Si-based transmission spectrum (black line) correspond to the magnetic dipolar, electric dipolar, and magnetic quadrupolar resonances, respectively. As for the Au NR, only electric dipolar resonance is excited, which is detrimental to produce a large $C_A$ on chiral medium positions.

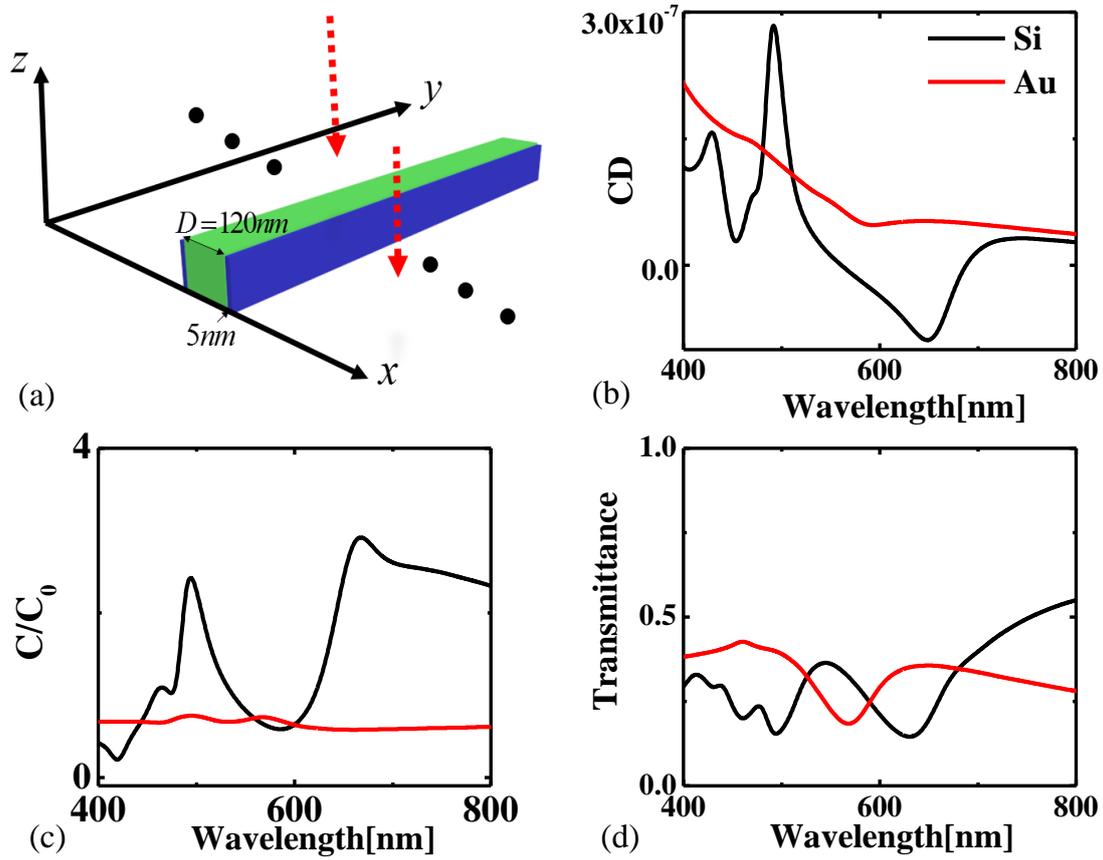

**Fig. 2** (a) shows the geometry and coordinate of the hybrid system, which consists of the periodic chiral medium stripe and NR. (b) shows the spectra for the enhanced CD based on the Si (black line) and Au (red line) NRs. (c) shows the enhancement of the optical chirality density as a function of the incident wavelength for the periodic NRs. (d) shows transmission spectra for the periodic NRs.

**Surface-enhanced CD spectra based on the gap nanoantenna systems**

In the above part, we concentrate our discussions on the surface-enhanced CD signal by the single nanoantenna systems. In this section, we will discuss the chiroptical amplification effect with the chiral molecule (medium) locating in the hotspots of the gap nanoantennas. Firstly, we consider the hybrid system, randomly oriented in the water solution, composed of a single chiral molecule and a NS dimer, shown in Fig. 3a. Here the parameters used for the chiral molecule and NSs are taken identically with the single NS systems. For comparisons, the corresponding results for the Au nanodimers (with the same size) are also presented.

Fig. 3b shows the CD signals induced by the Si and Au NS dimers with the separation distance between two NSs being d=5 nm. We can see that the value of the CD peak based on the Si NS dimer (black line) is still two times larger than that of the Au counterpart (red line), although the electric field is extremely strong in the gap of the Au NS dimer. In Fig. 3c, we plot the enhancements of the optical chirality density on chiral molecular position induced by Si (black line) and Au (red line) nanodimers with the wave vector of the incident left-hand circular polarized light being along y-axis. It is clearly shown that the Si NS dimer can amplify the local electromagnetic density of chirality more effectively than the Au NS dimer. In Fig. 3d, the far-field extinction efficiencies of the Si and Au nanodimers are presented. We can see that the Si-based nanodimer sustains both electric and magnetic resonances. This is beneficial for enlarging the electromagnetic density of chirality on the position of chiral molecules.

In Fig. 3e and 3f, we calculate the CD signals for the Si and Au NS dimers with different separation distances, respectively. We can see that the CD signals induced by Si NS dimers are always much larger than that of Au counterparts with different separation distances in a wide frequency range.

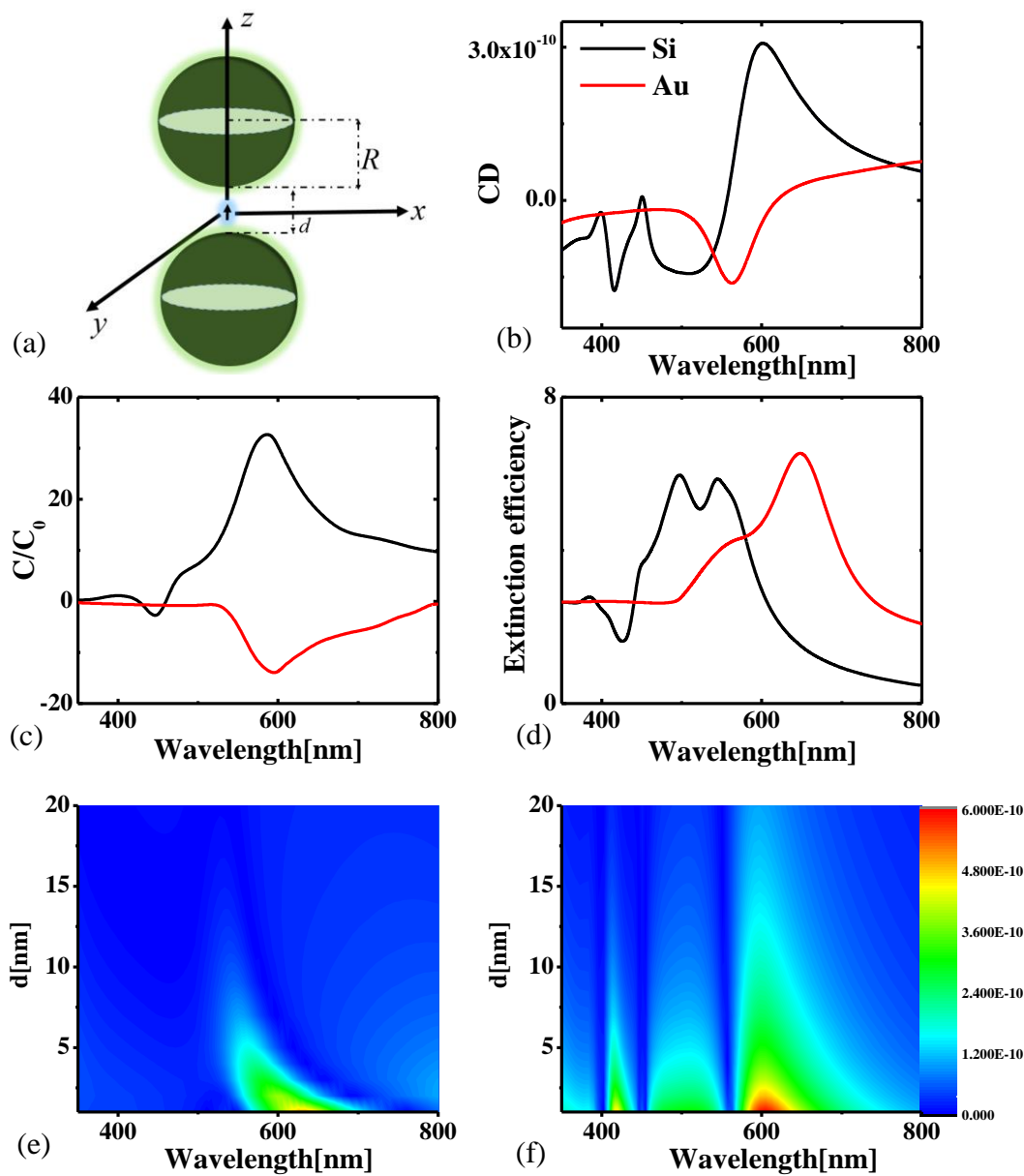

**Fig. 3** (a) shows the geometry and coordinate of the hybrid system, which consists of a chiral molecule and a NS dimer. The chiral molecule is put at the origin of the coordinate. The radius of the NS is R. The separation distance between two NSs is d. (b) shows the spectra for the enhanced CD based on the Si (black line) and Au (red line) NS dimers. The radius of the nanoparticle is $R$=65 nm. The separation distance of the nanodimer is d=5 nm. (c) shows the enhancement of the electromagnetic density of chirality as a function of the incident wavelength for the NS dimers. (d) shows the far-field extinction efficiency for the NS dimers. (e) and (f) display the relationship between the separation distances of the Au and Si nanodimers and the absolute values of the corresponding CD signals.

As shown from the above results, the Si nanodimer can induce a larger CD signal than the Au nanodimer with only one chiral molecule located in the nanogap. In the following, we will discuss the system that the chiral medium stripes are located in the hotspots of the periodic gap NR with the incident wave vector along z-axis. The corresponding schematic is shown in Fig. 4a. The dimensions of the NRs are the same to the models described in Fig. 2. And the lengths along x- and z-axis of the chiral medium strips are 2 nm and 120 nm, respectively. The gap size (x direction) is 5 nm, and the period is chosen to be 500 nm (the influences of higher diffraction orders on the CD signal are negligible).

The CD peak induced by the periodic Si gap NR is nearly five times larger than that of the corresponding Au counterpart, shown in Fig. 4b. In Fig. 4c, we plot the averaged enhancements of the optical chirality density on the position of chiral medium stripes induced by the Si (black line) and Au (red line) gap NRs. The averaged enhancement factor for the optical chirality density induced by the Si gap NR is nearly fifteen times larger than that by the Au counterpart. This enlarged difference of $C_A$ (comparing with the case of the single chiral molecule, Fig. 3c) results from the opposite handedness of the chiral near fields induced by the plasmonic nanostructures on the chiral medium stripe.[26] Fig. 4d presents the transmission spectra for the periodic Si (black line) and Au (red line) gap NRs. We can see that both electric and magnetic modes (including dipolar and higher-order modes) are excited in the Si-based gap NR. While, only one electric mode is excited in the periodic Au gap NR. Consequently, the Si gap NR can produce a larger $C_A$ than the Au counterpart, which is beneficial for the surface-enhanced CD spectroscopy.

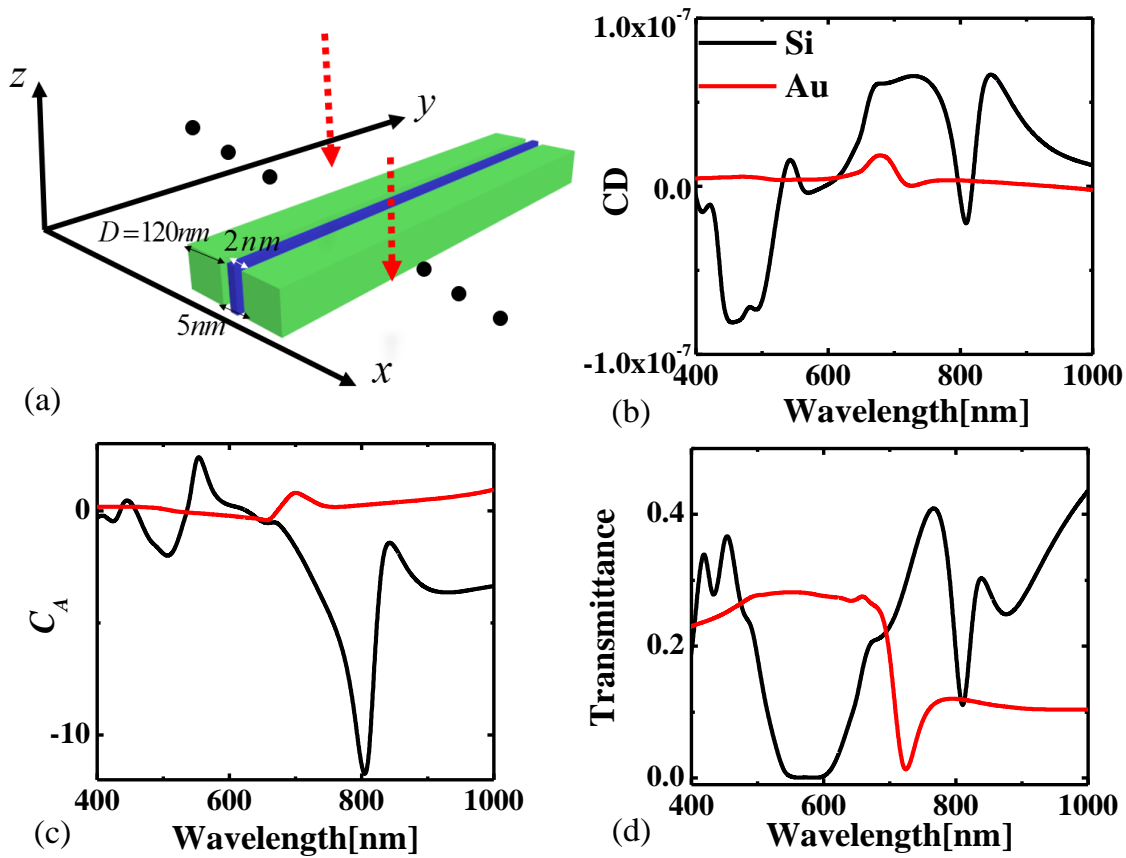

**Fig. 4.** (a) shows the geometry and coordinate of the hybrid system, which consists of the periodic chiral medium stripe and gap NR. (b) shows the spectra for the enhanced CD based on the Si (black line) and Au (red line) gap NRs. (c) shows the averaged enhancement of the electromagnetic density of chirality as a function of the incident wavelength for the gap NRs. (d) shows the transmission spectra for the gap NRs.

**Temperature effect**

From the above results we can see that both the single Si nanoparticle and nanodimer can produce much larger CD signals than the Au counterparts. This is due to the simultaneous electric and magnetic resonances in the Si nanostructures. On the other hand, the low intrinsic losses of the Si nanostructures in the optical and near-infrared regimes can open possibilities to overcome heat generation issues during the process of CD measurement. In this section, we will concentrate on the temperature effects of the systems, which stem from the absorption of incident radiation by Au and Si nanostructures. We employ a simple steady state model and assume that the thermal dissipation rate of the surrounding matrix is large relative to the local heating rate in the following simulations. The whole process of light absorption and subsequent heat transfer between the nanostructures and the surrounding medium have been modeled by means of finite element methods (Comsol Multiphysics).[27, 28, 67]

Fig. 5a shows the equilibrium distribution of the temperature increment around the single NS with radius R=65nm. The upper and lower panels correspond to Si and Au NSs, respectively. The wavelengths of the incident left-hand circular polarized light are tuned to be 580nm for Si and 540nm for Au, which correspond to the maxima of CD signals. The power densities of the incident light for are chosen to be $5 mW/um^2$. As expected, the increased temperature around the Au NS exceeds 200 K, which is nearly five times larger than the increased temperature of the Si NS. Fig. 5b shows the increased temperature on the molecular position around the Si (black line) and Au (red line) NSs as a function of the wavelength of the incident left-hand circular polarized light. The black and red arrows mark the locations of the Si and Au NSs induced CD peaks. Obviously, the Si NS is a better nanoantenna with ultralow heat to enhance the CD signal. We also analyze the temperature effect of the nanodimer systems. As shown in Fig. 5c, we plot the distributions of the increased temperature around the Si (upper panel) and Au (lower panel) NS dimer systems with radius R=65nm and separation distances d=5nm. The wavelengths of the incident left-hand light are chosen to be 605nm for Si and 563nm for Au, which correspond to the wavelengths with the maxima of CD signals. Remarkably, the increased temperature around the Si NS dimer is much less than that of the Au NS dimer. Fig. 5d shows the increased temperature on the molecular position around the Si (black line) and Au (red line) NS dimers as a function of the wavelength of the incident left-hand circular polarized light. The black and red arrows, in Fig. 5d, mark the positions of the CD peaks induced by Si and Au nanodimers, respectively. It is clearly shown that the increased temperature at the CD peak induced by the Si nanodimers is only 25 K, which is far less than the temperature increment, 240 K, around the Au nanodimers.

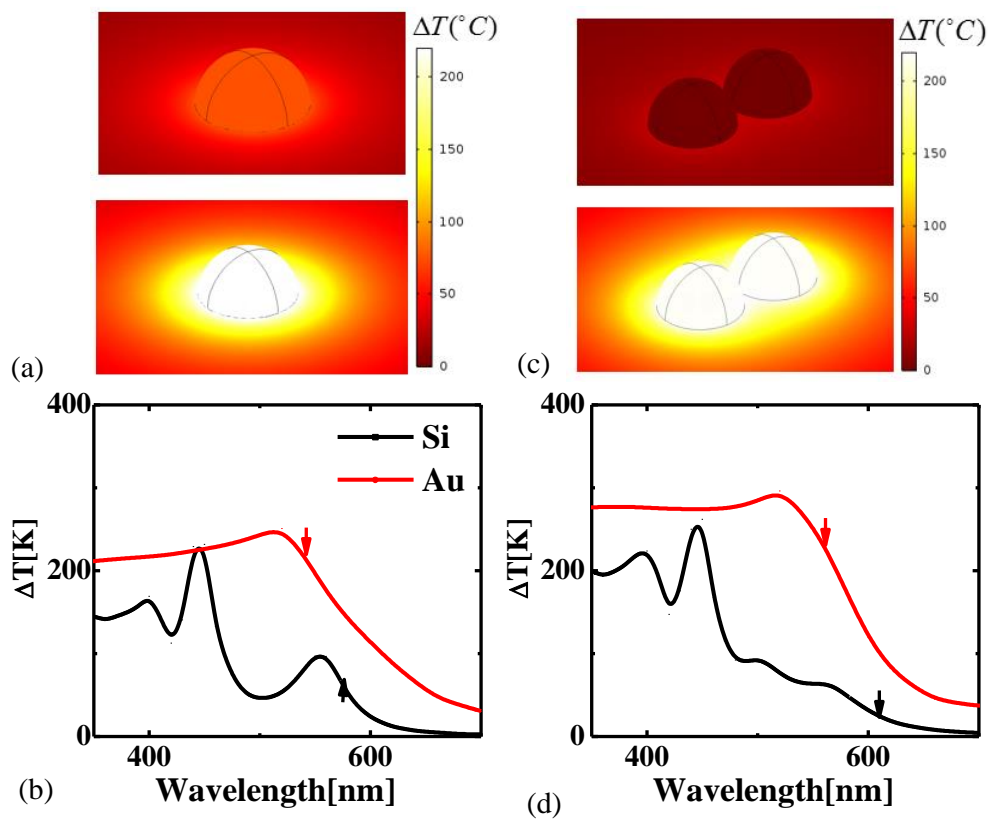

**Fig. 5** (a) Equilibrium distributions of temperature increment around the Si (upper panel, 580nm) and Au (lower panel, 540nm) NSs. (b) Temperature increment as a function of the wavelength of the incident left-hand circular polarized light for the single NS systems. (c) Equilibrium distributions of temperature increment around the Si (upper panel, 605nm) and Au (lower panel, 563nm) NS dimers. (d) Temperature increment as a function of the wavelength of the incident left-hand circular polarized light for the NS dimer system. The incident light intensity is $5mW/um^2$.

From the above results, we can see that Si-based nanoantenna can maintain the temperature increment lower than 50K at wavelengths with maxima CD signals. Most of biologically active molecules will not undergo the structure transition on this occasion. However, the heat produced by Au-based nanoparticles is nearly ten times larger than the Si counterparts. In such a high temperature environment, many molecules will undergo the irreversible transformation. It will destroy the molecular chirality and result in a failure of chirality charaterization.

**Conclusion**

In conclusion, we have demonstrated numerically that the molecular CD signal can be strongly enhanced based on high-index dielectric nanostructures. Due to the existence of both the electric and magnetic resonances in Si nanoantennas, the much stronger near-field chiral interactions between molecules and antennas can be produced. Taking plasmonic nanostructures as comparison, we find that the CD signals enhanced by the Si nanostructures are always much larger than that of Au counterparts over a wide range of the sizes. It is worth noting that sizes used for the Au and Si dimer nanoantennas are not optimal to enhance the molecular CD signal. Actually, the radius nearly equals to 30nm is the ideal condition. In this case, the Si-based CD peak is also larger than the Au-based counterpart. However, the corresponding optical absorption is more tremendous and will result in a more significant temperature increment (see in supporting information). In particular, due to the low absorption loss in dielectric systems, the silicon-based nanoantennas generate a negligible temperature increment in the nanoparticles and surrounding medium. This is very beneficial for detecting the enhanced CD signals.

## Acknowledgements

This work was supported by the National Key Basic Research Special Foundation of China under Grant No. 2013CB632704 and the National Natural Science Foundation of China through Grants No. 61421001 and 11574030.